\documentclass{emulateapj}
\shortauthors{Sommer-Larsen \& Toft}
\shorttitle{Simulating Compact Galaxies}
\begin{document}

\title{Cosmological Simulations of Massive Compact High-z Galaxies}
\author{J. Sommer-Larsen\altaffilmark{1, 2, 3} \& S. Toft\altaffilmark{1} }

\altaffiltext{1}
{Dark Cosmology Centre, Niels Bohr Institute, University of Copenhagen, Juliane Maries Vej 30, DK-2100 Copenhagen, Denmark, email: jslarsen@astro.ku.dk (JSL), sune@dark-cosmology.dk (ST)}

\altaffiltext{2}
{Excellence Cluster Universe, Technische Universit\"at M\"unchen,
  Boltzmannstrasse 2, 85748 Garching, Germany}

\altaffiltext{3}
{Marie Kruses Skole, Stavnsholtvej 29-31, DK-3520 Farum, Denmark}

\begin{abstract}

In order to investigate the structure and dynamics of the recently
discovered massive ($M_*\ga10^{11} M_\odot$) compact
$z\sim2$ galaxies, cosmological hydrodynamical/N-body simulations of
a $\sim50000$ Mpc$^3$ co-moving (lagrangian), proto-cluster region
have been undertaken. At $z=2$, the highest resolution simulation 
contains $\sim5800$ resolved galaxies, of which 509, 27 and 5 have 
$M_*>10^{10} M_\odot$, $M_*>10^{11} M_\odot$ and $M_*>4\times10^{11} M_\odot$,
respectively.   
Total stellar masses, effective radii and characteristic stellar
densities have been determined for all galaxies. At $z=2$, for the 
definitely well resolved mass range of $M_*\ga10^{11} M_\odot$, we 
fit the relation 
$R_{\rm{eff}}=R_{\rm{eff},12}M_{*,12}^{1/3}$ to the data, 
where $M_{*,12}$ is the total 
stellar mass in units of $10^{12} M_\odot$. 
This yields $R_{\rm{eff},12}=(1.20\pm0.04)$~kpc, in  
line with observational findings for compact $z\sim2$ galaxies, 
though somewhat {\it more} compact
than the observed average. The only line-of-sight velocity dispersion
measured for a $z\sim2$ compact galaxy is very large, 
$\sigma_{*,p}=510^{+165}_{-95}$ km/s \citep{vD.09}.   This value can
be matched at about the 1-$\sigma$ level, although a somewhat
larger mass than the estimated $M_*\simeq2\times10^{11} M_\odot$ is indicated.
For the above mass range, the galaxies have an average axial ratio 
$<b/a> = 0.64\pm0.02$
with a dispersion of 0.1, and an average rotation to 1D velocity 
dispersion ratio $<v/\sigma> = 0.46\pm0.06$ with a dispersion
of 0.3, and a maximum value of $v/\sigma \simeq 1.1$. 
Rotation and velocity anisotropy both contribute significantly in 
flattening the compact galaxies. Some of the observed compact galaxies
appear flatter than any of the simulated galaxies. 
Finally, it is found that the massive compact galaxies are strongly baryon 
dominated
in their inner parts, with typical dark matter mass fractions of order
only 20\% inside of $r=2R_{\rm{eff}}$. 

\end{abstract}

\vspace{2 cm}

\section{Introduction}
Observations indicate that a considerable part of 
the massive spheroidal stellar systems observed locally are already present 
at redshift 
$z \ga 2$, but that only a small fraction of these galaxies are fully assembled \citep{vD.08}.   
Typically, the proto-spheroidals are smaller by a factor of 3-6 
compared to present-day ellipticals at similar masses \citep{2005ApJ...626..680D,2007MNRAS.374..614L,Z.07,2007ApJ...671..285T,
2007MNRAS.382..109T,vD.08,F.08,B.08,2008A&A...482...21C,2008arXiv0810.2795S, W.09,T.09}. 
The stellar mass densities of the galaxies (inside of the effective radius) are at least one order of magnitude higher \citep{vD.08,2009ApJ...695..101D}, and their surface brightness is correspondingly significantly higher 
relative to low redshift galaxies of similar mass \citep{2008A&A...482...21C, 2009ApJ...695..101D, T.09}.  

The observed properties of $z\sim2$ proto-elliptical galaxies are hard to 
reconcile with some currently proposed (idealized) schemes for the formation  
of elliptical galaxies: the observations indicate 
that most early-type galaxies did not form in a simple monolithic 
collapse or a binary merger of gas-rich disks at high redshift, unless their 
increase in size is due to subsequent considerable (and unlikely)
amounts of mass loss \citep{2009ApJ...695..101D, 2008ApJ...689L.101F}.
Moreover, simple passive evolution of the stellar populations is in 
contradiction with observations of local ellipticals 
\citep{vD.08}.

Dry (i.e. gas-poor, collisionless) minor mergers and stellar accretion events  
are the prime candidates for the strong size evolution of massive stellar 
spheroids at $z\la2$  \citep{2006ApJ...636L..81N,2006ApJ...648L..21K,
2006ApJ...640..241B,2006ApJ...652..270B,2008ApJ...688..789G,
2009MNRAS.397..506K,2009arXiv0902.0373R,2009ApJ...691.1424H,
2009arXiv0903.4857V,2009ApJ...697.1369B,N.09, B.09} 
as the additional presence of a dissipative component in a major merger event 
would limit the size increase (see e.g. \citealp{2007ApJ...658...65C}). The observed ellipticals are already very massive 
at high redshift, thus we expect, e.g,. from the shape of the mass function, 
that minor mergers should be much more common than major mergers until $z=0$
\citep{2006ApJ...640..241B,2008MNRAS.388.1537M,
2008ApJ...688..789G,2009MNRAS.397..506K,2009ApJ...697.1369B}\\

The compact $z\sim2$ galaxies observed by, e.g., \cite{vD.08} are very
massive, with stellar masses of up to $3-4\times10^{11} M_\odot$. Moreover,
in the one case where the stellar line-of-sight velocity dispersion has
been measured, this is also very large, $\sigma_{*,p}=510^{+165}_{-95}$ km/s
(galaxy 1255-0; \citealp{vD.09}). From a theoretical/numerical point of
view, an obvious question is whether
$z\sim2$ galaxies of such extreme characteristics can be reproduced by
current cosmological state-of-the-art galaxy formation simulations. 
In a recent work, \cite{N.09} used a very high resolution simulation
of the formation of an individual galaxy to show that a galaxy with
$z\sim2$ characteristics similar to the observed compact galaxies can be
formed in conventional $\Lambda$CDM hydrodynamical/gravitational
simulations. The galaxy has, at $z\sim2$, an effective radius of 
$R_{\rm{eff}}=0.7\pm0.2$~kpc, and a stellar mass of 
$M_*\simeq 7\times10^{10} M_\odot$, at the lower end of the observed
mass range --- note, that the fact that less massive compact $z\sim2$
galaxies have not been spectroscopically confirmed so far, 
most likely is simply an observational limitation.\\

In order to build up a statistical sample of $z\sim2$, potentially compact
galaxies, and to span the entire observed mass range, we study in this 
paper the formation of galaxies in a $\sim50000$ Mpc$^3$ co-moving
(lagrangian), proto-cluster region.  
Such regions are some of the first regions in which the
growth of small density fluctuations goes non-linear, later causing
re-collapse and galaxy formation. They will hence potentially host
some of the most massive galaxies in the $z\sim2$ Universe. 
 

\begin{deluxetable*}{l c c c c c c c c}
\tablecolumns{9}
\tablewidth{0pc}
\tablecomments{Numerical  
characteristics of the simulations: mass of gas/star/DM particles and the respective gravitational softening
lenghts; total number of particles and initial redshift of run.}
\tablehead{
\colhead{run} &  \colhead{$m_{gas}$} &  \colhead{$m_{*}$} & \colhead{$m_{DM}$} & \colhead{$\epsilon_{gas}$}  &  \colhead{$\epsilon_{*}$}  &  \colhead{$\epsilon_{DM}$}  &
\colhead{$N_{tot}$}  &  \colhead{$z_i$} \\
        &  &\colhead{ [$10^7 M_{\odot}/h$]} &  &  & \colhead{[kpc/$h$]} &  &  & 
}
\startdata
Coma1      & 3.9 & 3.9 & 22 & 0.70 & 0.70 & 1.24 & 7100000 & 39\\
Coma2      & 3.9 & 3.9 & 22 & 0.35 & 0.35 & 0.62 & 7100000 & 39\\
Coma3      & 3.9 & 3.9 & 22 & 0.17 & 0.17 & 0.31 & 7100000 & 39\\
Coma3noSF  & 3.9 & 3.9 & 22 & 0.17 & 0.17 & 0.31 & 7100000 & 3\\
E4         & 0.073 & 0.073 & 0.42 & 0.20 & 0.20 & 0.36 & 1300000 & 39\\
E4SS       & 0.073 & 0.073 & 0.42 & 0.10 & 0.10 & 0.18 & 1300000 & 39\\
E4SSnosf   & 0.073 & 0.073 & 0.42 & 0.10 & 0.10 & 0.18 & 1300000 & 3\\
E4HR       & 0.0091 & 0.0091 & 0.052 & 0.10 & 0.10 & 0.18 & 9800000 & 59\\
E4HRnoSF   & 0.0091 & 0.0091 & 0.052 & 0.10 & 0.10 & 0.18 & 9800000 & 3\\
\enddata
\label{tab:data}
\end{deluxetable*}

Romeo et~al.~(2005, 2006), \cite{D.05} and \cite{SL.05} presented fully
cosmological simulations of galaxy groups and clusters. The TreeSPH
code used was building on the code used for simulating galaxy formation 
\citep[e.g.,][]{SL.03}, improved to include modeling of non-instantaneous 
chemical evolution \citep{L.02}, metallicity-dependent, atomic radiative 
cooling, strong supernova, and (optionally) AGN, driven galactic winds 
and thermal conduction. The two clusters simulated have $z$=0 virial
masses $M_{vir}\sim3\times10^{14}$ and $1.2\times10^{15} M_\odot$,
one approximately the size of the Virgo cluster and the other of the 
Coma cluster. They were both selected
to be fairly relaxed, and both display central prominent cD
galaxies at $z$=0.

In this paper we re-simulate the ``Coma'' proto-cluster region
at higher mass resolution and vastly higher force resolution than
used in the above works. To this end, we use a conventional version
of the hydro/gravity TreeSPH code GADGET-2 \citep{2005MNRAS.364.1105S}. 
The version used does
not include chemical evolution and metallicity dependent radiative
cooling, and also not radiative transfer of the meta-galactic UV field.
Nevertheless, it is sufficient for the purposes of this paper, which
is to study the basic structural and dynamical characteristics of very
massive, $z\sim2$ galaxies. A more sophisticated modeling, based on
the code described in \cite{R.05} etc., will be
presented in forthcoming papers. 

In order to address issues pertaining to the mass resolution of the
proto-cluster simulation, we supplement the simulations with
high and ultra-high resolution simulations of a proto-elliptical galaxy
region. The largest galaxy forming in this region is, at $z\sim0$, an
elliptical galaxy of stellar mass $M_*\simeq1.0\times10^{11} M_\odot$,
a bit less than that of the galaxy studied by \cite{N.09} at similar
mass and force resolution.\\  

This paper is organized as follows:
the code and the simulations are described in section 2, the results
obtained are presented in section 3 and discussed in section 4, and, 
finally, section 5 constitutes the conclusion.

Throughout the paper a
$\Lambda$CDM cosmology with $\Omega_{\rm{M}} = 0.3, \
\Omega_\Lambda = 0.7$ and a Hubble constant \textsc{H}$_0 = 70$
km\,s$^{-1}$ Mpc$^{-1}$ is assumed. 

\section{The code and simulations}
The simulations were undertaken with the hydro/gravity TreeSPH code GADGET-2 
\citep{2005MNRAS.364.1105S}.  

Star formation and feedback from supernovae was included using 
the sub-grid multiphase model of \cite{2003MNRAS.339..289S}. An 
over-density contrast of $\Delta > 55.7$ is required for the onset of star formation 
to avoid spurious star formation at high redshift. The threshold hydrogen number 
density for star formation is $n_{\rm{H,thresh}} = 0.13 \ \rm cm^{-3}$ and the star formation time-scale at the threshold density is $t_{*,{\rm thresh}} = 
2.2$~Gyr (in general, the star formation time-scale above the threshold density
is $t_* = t_{*,{\rm thresh}} \sqrt{n_{\rm{H,thresh}}/n_{\rm{H}}}$).

Radiative cooling and heating 
was invoked using a primordial cooling function and a uniform UV background 
(UVB) radiation field peaking at at $z\simeq 2-3$ \citep{HM96}. No radiative 
transfer (RT) of the UVB was performed - for the very massive galaxies
targeted in this work, RT effects on the dynamics of galaxy formation
are expected to be negligible.  

In calculating the supernova feedback, a Salpeter~(1955) stellar initial
mass function (IMF) is assumed. As this is the only way the simulations
depend on the choice of IMF (we only present results in terms of stellar
{\it masses}, and chemical evolution is not invoked), the results presented
are essentially IMF independent (e.g., Springel \& Hernquist 2003).

At $z$=0, part of the Coma simulation
volume ends up as a galaxy cluster of virial mass 
1.2x10$^{15}$ M$_{\odot}$ and X-ray emission weighted temperature
6.0 keV.  The proto-cluster region was selected from a
cosmological, DM-only simulation of a flat $\Lambda$CDM model, with
$\Omega_M$=0.3, $\Omega_b$=0.045, $h$=0.7 and $\sigma_8$=0.9 and a box-length
of 150 $h^{-1}$Mpc. 
Mass and force resolution was increased in, and gas particles added to,
the Lagrangian proto-cluster region. Using GADGET-2, the region was then
re-simulated using 7.1 million baryonic+DM particles with $m_{\rm{gas}}$=$m_*$=
3.9x10$^7$ and $m_{\rm{DM}}$=2.2x10$^8$ $h^{-1}$M$_{\odot}$ for the high 
resolution gas, star and dark matter particles. 

As the galaxies we want to simulate are very compact at $z\sim 2$,
with $R_{\rm{eff}}\sim 1$~kpc, it is critical that the force resolution is
high. To assess the numerical effects of force resolution, we carried out 
the (otherwise identical) proto-cluster region simulations at three different 
force resolutions, with gravity softening lengths of
($\epsilon_{\rm{gas}}$,$\epsilon_*$,$\epsilon_{\rm{DM}}$)=
(0.70,0.70,1.24), (0.35,0.35,0.62) and (0.17,0.17,0.31) $h^{-1}$kpc, 
respectively. Moreover, hydrodynamical smoothing lengths were restricted 
to be not smaller than 10\% of the gas and star particle gravity softening 
lengths. As will be shown in the next section, only for the last set
of gravity softening lengths is a realistic modeling of the $z\sim 2$
compact galaxies achieved. In addition, it is shown that the compact
galaxies are strongly baryon dominated, so what matters for the resolution
of the galaxies is the force resolution of the gas and stars. 

In order to assess also effects of mass resolution, we ran an additional
set of simulations of the formation of an individual galaxy, known 
from previous work \citep{SL.03} to become a $M_*\sim 10^{11} M_\odot$ 
elliptical galaxy at $z\sim0$ (we shall dub this galaxy ``E4'').
Three simulations of this individual galaxy were carried out:
One, (``E4HR''), with 9.8 million particles, $m_{\rm{gas}}$=$m_*$=
9.1x10$^4$ and $m_{\rm{DM}}$=5.2x10$^5$ $h^{-1}$M$_{\odot}$ and
($\epsilon_{\rm{gas}}$,$\epsilon_*$,$\epsilon_{\rm{DM}}$)= 
(0.10,0.10,0.18) $h^{-1}$kpc, and two, (``E4'' and ``E4SS''), 
with 1.3 million particles,
$m_{\rm{gas}}$=$m_*$=7.3x10$^5$ and $m_{\rm{DM}}$=4.2x10$^6$ 
$h^{-1}$M$_{\odot}$, and
($\epsilon_{\rm{gas}}$,$\epsilon_*$,$\epsilon_{\rm{DM}}$)= 
(0.20,0.20,0.36) and (0.10,0.10,0.18) $h^{-1}$kpc, respectively.

For all simulations, gravitational softening lengths were fixed in co-moving
coordinates till $z$=6, subsequently in physical coordinates.
Numerical parameters of the simulations are summarized in 
Table.~\ref{tab:data} 

\subsection{Simulations with no ``late'' star formation}
Observationally, the compact $z\sim2$ galaxies in general appear
to contain little stars of ages $\la 1$ Gyr, corresponding to
formation redshifts less than about three. The reason for this
truncation of star formation, if real, is not known, but is likely
related to the effect of accretion onto super-massive black holes at
the centers of the galaxies, and the resulting violent feedback effects
\citep[e.g.,][]{B.06,C.06}.

In the simulations
described above, AGN feedback was not invoked, and some residual star 
formation takes place also at $z\la 3$. To assess, in a simple way, 
how the results obtained depend on this
``late'', possibly spurious, star formation, we carried out three
additional simulations, one for the Coma proto-cluster region, and
two for the E4 proto-elliptical region. In each, star-formation and radiative cooling and heating was switched off at $z=3$, and the simulations were then 
continued to $z=0$. These simulations represent the extreme case of
purely ``dry merging'' and ``passive'' evolution since $z=3$ --- see also \cite{SLL09}.
Numerical parameters of these three additional simulations, dubbed
``Coma3noSF'', ``E4SSnoSF'' and ``E4HRnoSF'' are also given in 
Table.~\ref{tab:data}.

\section{Results}
\subsection{Identification of galaxies}
Galaxies, as represented by their stellar content, were identified
using the approach outlined in \cite{SL.05} and Romeo et~al. (2005, 2006).
In particular, care was taken to remove unbound stars from the 
galaxies --- this is mostly of relevance for the Coma simulations at
$z\sim0$, as the final cluster contains a large number of intra-cluster
stars \citep{SL.05}. 

Second, galaxies in process of merging were removed from the sample.
Specifically, at any $z$, galaxies containing multiple structures
separated by less than 10 kpc were removed. The fraction of such
galaxies was at any $z$ small, $f_{\rm{merging}}\la0.1$.

For the Coma runs, at $z\sim2$ the final galaxy sample comprises
$\sim5000$ galaxies down to the resolution limit of $M_*\sim5\times10^8
M_\odot$, corresponding to $M_{\rm{tot}}\sim5\times10^9 M_\odot$.
At $z\sim0$, due to the substantial amount of merging and tidal
destruction taking place from redshift 2 to 0, the number of identified
is reduced to about 2500. For the simulation with no star formation
since $z=3$, the effect is more pronounced: a reduction from about 3000
galaxies at $z=2$ to 1100 at $z=0$.

\subsection{Determination of galaxy effective radii}
As a first step in determining the effective radii of the galaxies,
for a given galaxy all (bound) stars within a radius of $r_0$ and
centered on the galaxy are selected. At $z\sim2$, a value of 
$r_0=10$ kpc is used. This is several times larger than the
``optical radius'' (see below), and the results presented in this
paper are in any case not sensitive to the exact choice of $r_0$. 
At $z=0$, $r_0=15$ kpc is used, except for the cD, for which
$r_0=50$ kpc is adopted. 

Next, stellar surface density profiles are determined as follows:
the spatial distribution of stars for a given galaxy is projected
along the three cardinal directions, and the resulting three
projected distributions co-added. The azimuthally averaged mass
surface density profile (including averaging over the three
cardinal directions) is subsequently obtained in annuli of
$\Delta R = 200$ pc, where $\Delta R$ is the difference between the
outer and inner radius of the annulus. 

At $z=0$, the ``optical
radius'', $R_{25}$, of the galaxy is then determined, by interpolation of the
surface density profile. $R_{25}$ is defined to correspond to
a mass surface density of
\begin{equation}
\Sigma_{M,25} = (M/L_B)_{25}\cdot\Sigma_{L_B,25} ~~,
\end{equation}
where $\Sigma_{L_B,25}$ is the surface brightness corresponding
to the optical radius of a galaxy, assumed here to be 
25 B-mag/arcsec$^2$, corresponding to 155 $L_{B,\odot}$/pc$^2$,
and $(M/L_B)_{25}$ is the projected B-band mass-to-light
ratio at $R_{25}$. Based on models incorporating full
chemical evolution, non-instantaneous super-nova driven energy
and heavy element feedback, UVB radiative transfer etc, and
adopting a Salpeter (1955) initial mass function we estimate
$(M/L_B)_{25}\sim5$ at $z$=0 (Sommer-Larsen \& Toft 2010), and
adopt this value in the following.

At $z\sim2$, it is less obvious how to determine the ``optical
radius'', $R_{\rm{opt}}$, of a galaxy. Motivated by \cite{H.09}, 
we define the
optical radius as the radius where the mass surface density is
$q$ magnitudes, {\it{i.e.}} a factor of 10$^{(q/2.5)}$, below the
central surface density, defined as the mean surface density
inside of the effective radius, $R_{\rm{eff}}$. The values of
$R_{\rm{opt}}$ and $R_{\rm{eff}}$ are determined through an
iterative procedure, as detailed below. A value of $q$=4 is
assumed --- we find that the results presented in this paper are insensitive
to moderate variations of $q$. 

Once the optical radius of the galaxy is determined, the total projected
stellar mass inside of $R_{\rm{opt}}$, $M_*$, is calculated. Next, the
effective radius, $R_{\rm{eff}}$, is determined as the projected
radius containing a projected stellar mass of $M_{*,1/2}=0.5M_*$ --- we are
in this assuming a constant mass-to-light ratio; this assumption
will be relaxed in Sommer-Larsen \& Toft (2010). At $z=0$, the 
determination of $R_{\rm{eff}}$ is straightforward; at $z\sim2$, 
$R_{\rm{opt}}$ depends on $R_{\rm{eff}}$, as described above, and
an iterative procedure is used in determining $R_{\rm{opt}}$ and 
$R_{\rm{eff}}$ --- this procedure, however, is fully robust, and also
quite straightforward. 

At $z=0$, it makes sense to compare effective radii determined in the
two different ways described above. For the more massive, well resolved
galaxies (see section 4), it is found that the effective
radii are consistent to within $\sim 5$\%.

\begin{figure}
\epsscale{1.2}
\plotone{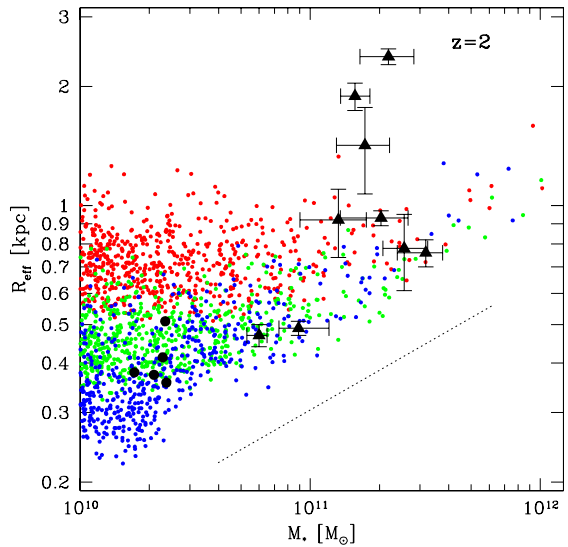}
\caption{Effective radii versus galaxy stellar masses at $z=2$, 
    for the proto-cluster simulations Coma1 (red dots), Coma2 (green dots) 
    and Coma3 (blue dots). Also shown are the results for five galaxies in
    the ultra-high resolution proto-elliptical simulation E4HR
    (large black dots). The statistical uncertainty on the data points
    is of order 5\% (from comparing results along the three different
    projection directions). Moreover is shown the 9 observational $z\sim2$
    compact galaxy data points from van Dokkum et~al.~(2008)
    (black triangles with error-bars). Finally is indicated a power-law
    of logarithmic slope 1/3 and arbitrary normalization, corresponding
    to $\rho_* = \rm{constant}$ (eq.[3]).
}
\label{fig:reffz2}
\end{figure}
\begin{figure}
\epsscale{1.2}
\plotone{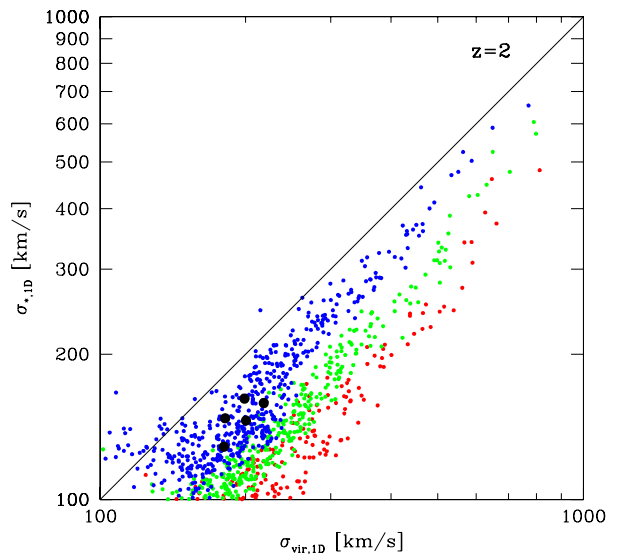}
 \caption{1D velocity dispersion versus 1D virial velocities for 
   galaxies of $M_*>10^{10} M_\odot$ in simulations 
   Coma1-3. Also shown are the results for five galaxies of 
   $M_*>10^{10} M_\odot$ for simulation E4HR. The colour coding is as in 
   Fig.1. The solid line indicates $\sigma_{*,1D}=\sigma_{vir,1D}$.}   
\label{fig:sigma}
\end{figure}
\subsection{Galaxy effective radii, velocity dispersions and 
stellar densities}
Figure~\ref{fig:reffz2} shows, at $z=2$, $R_{\rm{eff}}$ vs. $M_*$
for the three Coma simulations of varying force resolution. At the
high mass end, $M_*\sim10^{12} M_\odot$, $R_{\rm{eff}}\ga\epsilon_*$
for all three simulations, and the results of the three runs
agree fairly well. At lower masses, $M_*\sim10^{10}-10^{11} M_\odot$,
the agreement is clearly less good. This is not surprising, since,
judging from the highest resolution simulation, for such galaxy
masses, $R_{\rm{eff}}\la\epsilon_*$ for the simulation of poorest
force resolution, and $R_{\rm{eff}}\sim\epsilon_*$ for the intermediate
resolution simulation. This strongly indicates that, in this mass
range, only the highest resolution simulation, if any, can be
used for modeling of the compact galaxies. In order to assess
whether the resolution of the latter simulation is sufficient, we
now determine the velocity dispersions of the galaxies. Specifically,
we calculate, for each galaxy, the 1D velocity dispersion of the
stars inside of physical radius $r$=2$R_{\rm{eff}}$, $\sigma_{*,1D}$, 
and compare this to what is expected from dynamics (assuming that
the galaxies are baryon dominated in the inner parts --- see section 4), 
{\it{viz}}, 
\begin{equation}
\sigma_{\rm{vir},1D} \simeq \sqrt{\frac{1}{3}} 
\sqrt{\frac{GM_*(r<2R_{\rm{eff}})}{2R_{\rm{eff}}}} ~~.
\end{equation}
For simplicity we assume that all stellar mass resides inside of
$r=2R_{\rm{eff}}$, which will somewhat overestimate the values of
$\sigma_{\rm{vir},1D}$ --- see below.             
Shown in Fig.~\ref{fig:sigma} is $\sigma_{*,1D}$ vs. 
$\sigma_{\rm{vir},1D}$ for the three Coma simulations at $z=2$, with only
galaxies of $M_*>10^{10} M_\odot$ shown. As can be seen from the figure,
at the highest spatial resolution, $\sigma_{*,1D} \simeq 
\sigma_{\rm{vir},1D}$, whereas at poorer spatial resolutions this is not 
the case. This indicates that the highest spatial resolution is
sufficient for the modeling presented, although the mass resolution,
and, in particular, the two-body relaxation time, are possible issues ---
this is further discussed in the next section. The fact that, for
Coma3, $\sigma_{*,1D}$ still lies $\sim10$\% below $\sigma_{\rm{vir},1D}$
could be taken as an indication that the force resolution is still not
adequate. However, shown also in the figure, are results, at $z=2$, 
for five galaxies 
of $M_*>10^{10} M_\odot$ from the E4HR run, which has higher force resolution
and much higher mass resolution than Coma3. The data points are seen
to fall perfectly on the relation inferred from Coma3, indicating that
the $\sim10$\% offset is more likely due to the assumption that all
stellar mass resides inside of $2R_{\rm{eff}}$.

Given the above findings, we shall in the following, when analyzing
the proto-cluster simulations, only discuss results of the highest force 
resolution simulation, Coma3.
\begin{figure}
\epsscale{1.2}
\plotone{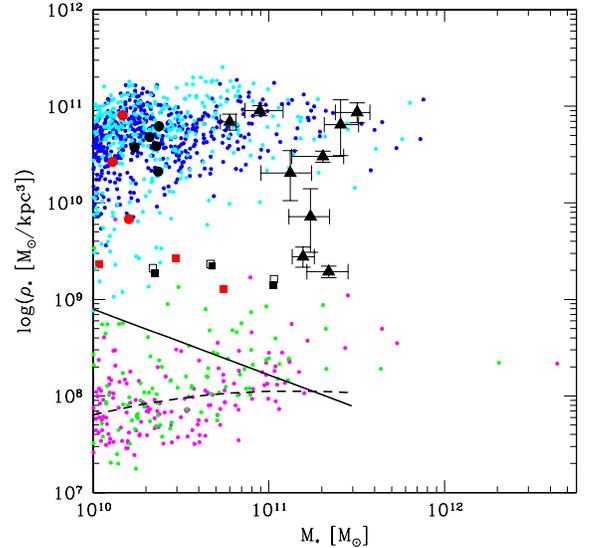}
 \caption{Characteristic stellar densities (eq.[3]) vs. galaxy
   stellar mass for galaxies at $z$=2 and 0. Results for the ``standard''
   proto-cluster simulation, Coma3, at $z=2$ are shown by blue dots 
   Moreover, shown by cyan dots, are the results
   for the proto-cluster simulation, Coma3noSF, where star-formation and 
   radiative cooling and heating has been switched off at $z=3$. 
   The corresponding $z=0$ results are shown by purple and green dots,
   respectively. Also,
   the results for the proto-elliptical region are shown: At $z=2$,
   only results for the ultra-high resolution simulations 
   E4HR and E4HRnoSF are shown 
   (by filled black and red circles, respectively) --- results for the other 
   three E4 runs are
   similar. At $z=0$ are shown, by filled black squares, results for
   simulation E4, by open black squares, results for E4SS, and, by filled red
   squares, results for E4SSnoSF. The statistical uncertainty on the data 
   points is of order 15\% (from comparing results along the three different
   projection directions). Finally, for comparison to observations, are 
   shown a) the 9 observational $z\sim2$
   massive compact galaxy data points from van Dokkum et~al.~(2008)
   (black triangles with error-bars), and b) the median loci of the
   SDSS $z\sim0$ galaxies from \cite{Sh.03}, corrected as described by \cite{2008A&A...482...21C}; early types (solid black curve), late types (dashed black curve).}
\label{fig:rhostar}
\end{figure}
\begin{figure}
\epsscale{1.2}
\plotone{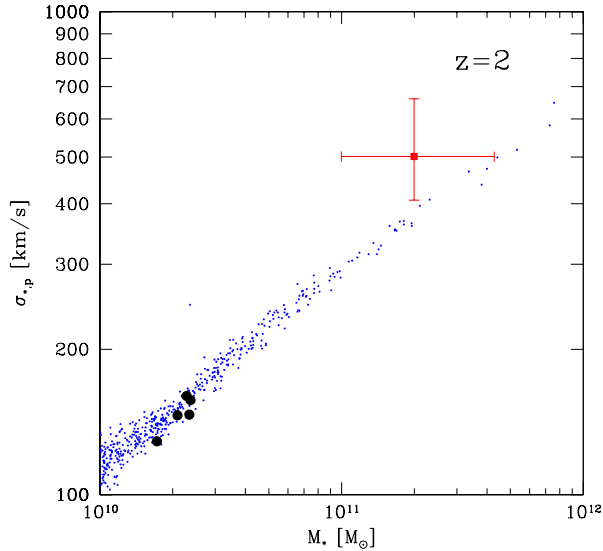}
 \caption{Projected velocity dispersion of galaxies in the Coma3
   proto-cluster simulation vs. galaxy stellar mass (blue dots). 
   Also shown are the results for the ultra-high resolution simulation
   E4HR (filled black circles). Finally is shown, by the filled
   square with error-bars, the measurement for the $z\sim2$ galaxy 1255-0.
   The horizontal error-bars indicate the potential mass range 
   \citep{vD.09,K.09} --- for more detail see text.}
\label{fig:sigp}
\end{figure}

Motivated by the finding below, that the characteristic stellar densities
of the galaxies are approximately independent of $M_*$, we fit a
relation of the form $R_{\rm{eff}} \propto M_*^{1/3}$ to the data
(to guide the eye, a line of logarithmic slope 1/3 is shown in
Fig.~\ref{fig:reffz2}).
Fitting to the Coma3 data in the definitely resolved mass range
of $M_*\ga10^{11} M_\odot$ (see section 4), we find 
$R_{\rm{eff}} = (1.20\pm0.04)M_{*,12}^{1/3}$ kpc,
where $M_{*,12}$ is the total stellar mass in units of $10^{12} M_\odot$.
The Coma2 simulation yields, for the same mass range, 
$R_{\rm{eff}} = (1.13\pm0.05)M_{*,12}^{1/3}$ 
kpc, indicating that the galaxies in this mass range are resolved.
For simulation Coma3noSF we find 
$R_{\rm{eff}} = (1.34\pm0.06)M_{*,12}^{1/3}$, so these galaxies are
only very marginally less compact than the Coma3 ones.
 
On the observational side,
for the 9 data points of \cite{vD.08}, the inferred stellar galaxy
masses depend on the stellar initial mass function (IMF) adopted.
For a Salpeter IMF we find, giving each data point
equal weight, $R_{\rm{eff}} = (1.53\pm0.28)M_{*,12}^{1/3}$ kpc; 
for a Kroupa IMF,
the case assumed in Figs. 1 and 3, the corresponding result is 
$R_{\rm{eff}} = (1.77\pm0.32)M_{*,12}^{1/3}$ kpc. Although the
Coma3 results are marginally consistent with the observations, when
assuming a Salpeter IMF, our results indicate that the proto-cluster
galaxies are {\it more} compact than the data of \cite{vD.08}.

In Fig.~\ref{fig:rhostar} we show, for the Coma3 and Coma3noSF
simulations, and at $z=2$ and $z=0$, galaxy characteristic
stellar densities, defined as
\begin{equation}
\rho_* = \frac{M_{*,1/2}}{\frac{4}{3} \pi R_{\rm{eff}}^3}  ~~,
\end{equation}
following \cite{vD.08}. Also shown are the $z\sim2$ compact galaxy data 
points from \cite{vD.08}. As can be seen from the figure, the results
for the two simulations are located similarly in the diagram, cf.
the discussion above. Moreover, the
simulations reproduce the upper locus of the observational
data quite well, but on average, the observational data fall somewhat
below the simulation. The reason for this, and the similar situation
in Fig.~\ref{fig:reffz2}, is likely that the observations cover a substantial
range of environments, whereas the simulations trace a specific
proto-cluster region. The dispersion in $\rho_*$ for the simulated
galaxies, at a given $M_*$, is about 60\%, much larger than the statistical 
uncertainty
of about 15\%, derived by comparing results along the three different
projection directions. This strongly indicates that the scatter seen
in the figure is real. The observational scatter is larger than that
of the simulated galaxies, specifically the scatter is a factor 
$\sim 2.5$ --- this again is most likely due to environmental effects.
The fact that the simulations can reproduce the properties of the most
compact, massive $z\sim2$ galaxies is the most important result of this
paper.

For comparison to present day galaxies is shown the median loci
of the $z\sim0$ SDSS galaxies, divided into early and late 
types --- see \citep{Sh.03} for details.
A thorough discussion of the morphology of the simulated galaxies
will be given in Sommer-Larsen \& Toft (2010), but here we just
note that the average gas fraction of the $z=0$ galaxies of 
$M_*>10^{10} M_\odot$ is $1.2 \times 10^{-3}$  and $1.0 \times 10^{-3}$ for the Coma3 and
Coma3noSF simulations, respectively. This strongly hints that
the bulk of the $z=0$ simulated galaxies have early type characteristics.  
At masses, $M_*\la10^{11} M_\odot$, $\rho_*$ of the simulated
galaxies falls somewhat below the locus of SDSS early type galaxies.
This may, at least partly, be a numerical effect, caused by two-body 
relaxation, as will be discussed in section 4.

\subsection{Projected velocity dispersions}
For comparison to observations we also calculate projected
(line-of-sight) stellar velocity dispersions for the galaxies at
$z=2$. Specifically, for a given galaxy we project the spatial stellar
distribution along the three cardinal axis, and calculate the
line-of-sight velocity dispersion, $\sigma_{*,p}$, for all stars
within projected distance $R=2R_{\rm{eff}}$ of the galactic
center, and averaging over the three projection directions. The
results are shown in Fig.~\ref{fig:sigp}, both for Coma3 and for
E4HR. Also shown is the observational result of \cite{vD.09} 
for the compact, $z=2.3$ galaxy 1255-0, $\sigma_{*,p}=510^{+165}_{-95}$
km/s. \cite{K.09} estimate the stellar mass of the galaxy to be
$M_*\sim 2\times10^{11} M_\odot$, and \cite{vD.09} the dynamical mass
to lie in the range $1.0-4.3\times10^{11} M_\odot$. Assuming a stellar
mass of $2\times10^{11} M_\odot$ the measurement of \cite{vD.09} is
consistent with the simulations at the 1.1-$\sigma$ level. But,
taken at face value, the simulations indicate a somewhat larger
galaxy mass, $M_*\sim 3-4\times10^{11} M_\odot$. We note that 1255-0,
with an effective radius of $R_{\rm{eff}}=0.78\pm0.17$ kpc, is one
of the most compact galaxies observed (for its stellar mass), and lies
close to the $R_{\rm{eff}}-M_*$ mean locus of the Coma3 galaxies.  
\begin{figure}
\epsscale{1.2}
\plotone{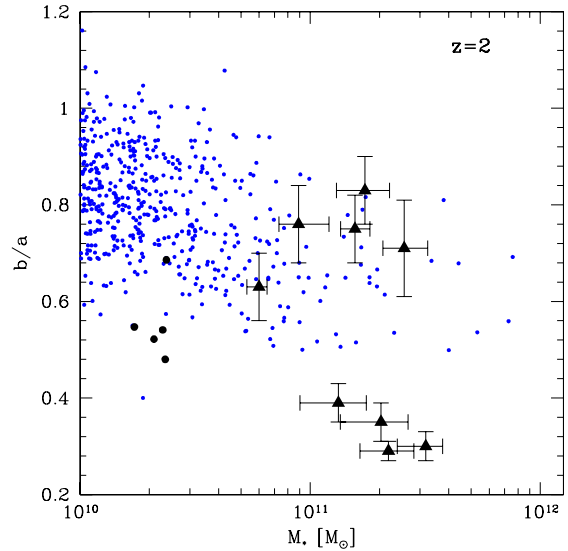}
 \caption{Axial ratios, as defined in eq.[4], of galaxies in the Coma3
   proto-cluster simulation vs. galaxy stellar mass (blue dots). 
   Also shown are the results for the ultra-high resolution simulation
   E4HR (filled black circles). Finally, is shown the 9 observational $z\sim2$
    compact galaxy data points from van Dokkum et~al.~(2008)
    (black triangles with error-bars). The E4HR galaxies appear on average
    flatter than the Coma3 galaxies of similar mass. This originates 
    likely, at least partly, in the ``low-mass'' Coma3 galaxies being affected by two-body relaxation effects, making them rounder, as discussed in section 4.
 }
\label{fig:axrat}
\end{figure}
\begin{figure}
\epsscale{1.2}
\plotone{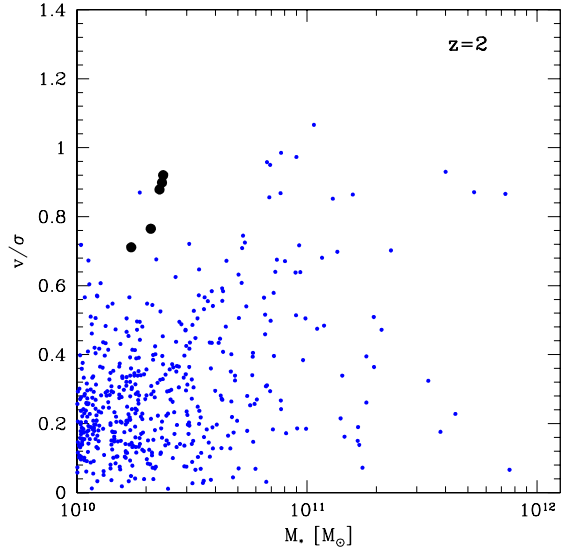}
 \caption{Ratio of rotation velocity to 1D velocity 
   dispersion ratio, $v/\sigma$, vs. galaxy stellar mass (blue dots). 
   Also shown are the results for the ultra-high resolution simulation
   E4HR (filled black circles) --- see text for more detail.}
\label{fig:vs}
\end{figure}
\begin{figure}
\epsscale{1.2}
\plotone{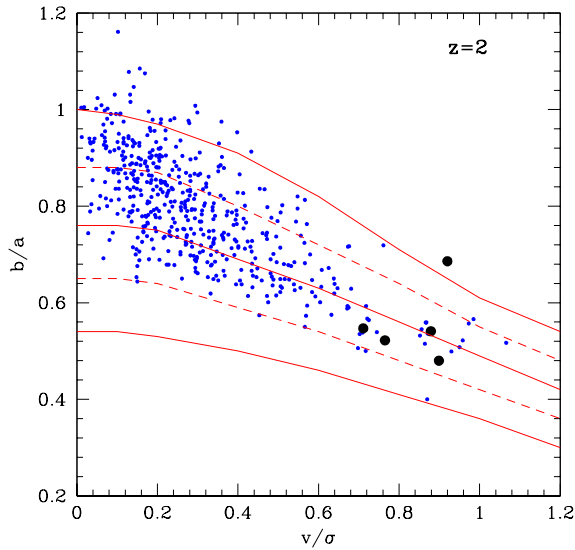}
 \caption{Axial ratios of galaxies in the Coma3 proto-cluster simulation vs. 
   $v/\sigma$ (blue dots). Also shown are the results for the 
   ultra-high resolution simulation E4HR (filled black circles). Finally, is 
   shown the theoretical predictions for for spheroids
   of concentric isodensity surfaces for velocity dispersion tensor
   anisotropy parameters (eq.[6]) $\delta$=0.0, 0.1, 0.2, 0.3 and 0.4
   (top down). One of the E4HR data points is characterized by a negative
   value of $\delta$ --- this galaxy has a disturbed spatial and
   kinematic structure, caused by a recent merging event.}
\label{fig:axrat_v}
\end{figure}

\subsection{Axial ratios, rotation and velocity anisotropy}
For each $z=2$ galaxy, the angular momentum of the stars inside
of $r=2R_{\rm{eff}}$ was determined, and the coordinates rotated
such that the new z-axis was aligned with the angular momentum
vector. The axial ratio is, for simplicity, determined as 
\begin{equation}
b/a = \frac{\sqrt{2}<|z|>}{<\sqrt{x^2+y^2}>} ~~, 
\end{equation}
which is exact for a prolate spheroid. The averaging is performed over all
stars inside of $r=2R_{\rm{eff}}$. In Fig.~\ref{fig:axrat} is
shown $b/a$ vs. $M_*$ for the Coma3 run at $z=2$. For the 27 galaxies
of $M_*>10^{11} M_\odot$, we find $<b/a> = 0.64\pm0.02$ with a dispersion 
of 0.1.

Next, the mean rotation velocity (about the z-axis) of each galaxy,
$v = <v_{\phi}>$, is determined. Moreover, the average 1D velocity dispersion 
along directions perpendicular to the z-axis, $\sigma$, is estimated as
\begin{equation}
\sigma = \sqrt{\frac{\sigma_R^2+\sigma_{\phi}^2}{2}} ~~, 
\end{equation}
where $R$ is the radial coordinate and $\phi$ is the azimuthal angle in
cylindrical coordinates. In Fig.~\ref{fig:vs} is shown the ratio
$v/\sigma$ vs. $M_*$. For the 27 galaxies
of $M_*>10^{11} M_\odot$, we find $<v/\sigma> = 0.46\pm0.06$ with a dispersion
of 0.3, and a maximum value of $v/\sigma \simeq 1.1$. Such rotation 
velocities are dynamically important, and result in flattened galaxies
even for isotropic velocity-dispersion tensors. To investigate whether
the galaxies are flattened by rotation only, we show in 
Fig.~\ref{fig:axrat_v} $b/a$ vs. $v/\sigma$. As can be seen, the two
quantities are correlated, so rotation clearly plays a role in 
shaping the galaxies. Also shown in the figure are curves of different 
\begin{equation}
\delta = 1 - \frac{\sigma_z^2}{\sigma^2} ~~, 
\end{equation}
calculated on the basis of the tensor virial theorem for spheroids
of concentric isodensity surfaces
\citep{BT87}. As can be seen, models with isotropic
velocity-dispersion tensors do not match the bulk of the galaxies
well; values of $\delta \sim 0.1-0.2$ are indicated.We conclude
that rotation and velocity anisotropy both contribute significantly in 
flattening the compact galaxies. 

\section{Discussion}
The results presented in the previous section indicate that the massive,
$z\sim2$ compact galaxies can be reproduced in the highest force resolution
simulations, but that the $z\sim0$ galaxies at masses $M_*\la10^{11} M_\odot$
get a bit too ``puffed up'' relative to observations. Given that the galaxies
are represented by relatively modest numbers of star particles 
($\la20000$ at $z\sim2$, and $\la150000$ at $z\sim0$; for the Coma3
galaxies, the number of star particles is related to the total stellar
mass by $N_* \sim 2\times10^4 M_{*,12}$, where $M_{*,12}$ is the total 
stellar mass in units of $10^{12} M_\odot$), and that
the $z\sim2$ galaxies are characterized by comparatively large stellar
densities, it is important estimate the two-body relaxation time
in the galaxies. Following \cite{BT87}, this can be expressed as
\begin{equation}
t_R \simeq \frac{N}{8~ln{\Lambda}} t_{\rm{cross}} ~~, 
\end{equation}
where $N$ is the number of particles, $ln{\Lambda}$ is the Coulomb
logarithm and $t_{\rm{cross}}$ the system crossing time. The 
Coulomb logarithm can for a system of gravitationally softened
particles be expressed as
\begin{equation}
ln{\Lambda} \simeq ln{\frac{r_{\rm{sys}}}{\epsilon_*}} ~~, 
\end{equation}
where $r_{\rm{sys}}$ is the radius of the system, and $\epsilon_*$ is
the gravitational softening length of the star particles (assumed to
dominate the mass density in the inner galaxy --- see below). 
The crossing time is given by 
\begin{equation}
t_{\rm{cross}} \simeq \frac{2r_{\rm{sys}}}{v_{\rm{cross}}} ~~, 
\end{equation}
where $v_{\rm{cross}}$ is the typical 1D velocity dispersion.
We shall conservatively assume that all stellar mass is contained
within $r_{\rm{sys}}=2R_{\rm{eff}}$. Moreover, based on Fig.\ref{fig:reffz2},
we assume the approximate relation 
$R_{\rm{eff}}\simeq0.35(M_*/10^{10}M_\odot)^{1/3}$ kpc. Finally,
we approximate the Coulomb logarithm as $ln{\Lambda} \simeq
ln{(1.4 \rm{kpc}/0.24 \rm{kpc})} \simeq 1.7$, assuming a typical
value of $R_{\rm{eff}} \simeq 0.7$ kpc, and inserting the gravitational
softening length of star particles in the highest force resolution
simulation. Setting
\begin{equation}
v_{\rm{cross}} \simeq \sqrt{\frac{1}{3}} \sqrt{\frac{GM_*}{2R_{\rm{eff}}}} ~~, 
\end{equation}
and inserting all the above in eq.[7], we obtain
\begin{equation}
t_R \simeq 18 ~\frac{M_{*,12}}{(m_*/5.6\times10^7 M_\odot)}~ \rm{Gyr} ~~, 
\end{equation}
where $M_{*,12}$ is the stellar mass of the galaxy in units of 
$10^{12} M_\odot$, and $m_*$ is the stellar particle mass.
Taken at face value, this indicates that for the present simulations
only galaxies of $M_*\ga5\times10^{11} M_\odot$ will be unaffected by
two-body relaxation during the $\sim10$ Gyrs evolution from $z\sim2$
to $z=0$. Note, however, that cold accretion and merging will
cause the galaxies to gradually ``puff up'' \citep[e.g.,][]{N.09}, which will
lessen the above constraint. Despite this, it is very likely that
the structure of the $z\sim0$ galaxies of $M_*\la10^{11} M_\odot$ has been affected
by two-body relaxation. In fact, at
$z=2$, the two-body relaxation time is comparable to the mean age of
the stellar population at $M_*\sim10^{11} M_\odot$, so even at
$z=2$, the structure of the simulated galaxies of $M_*\la10^{11} M_\odot$ 
may in principle be somewhat affected by two-body relaxation effects.
We shall hence conservatively denote the mass range
$M_*\ga10^{11} M_\odot$, at $z=2$, the definitely well resolved mass
range.
\begin{figure}
\epsscale{1.2}
\plotone{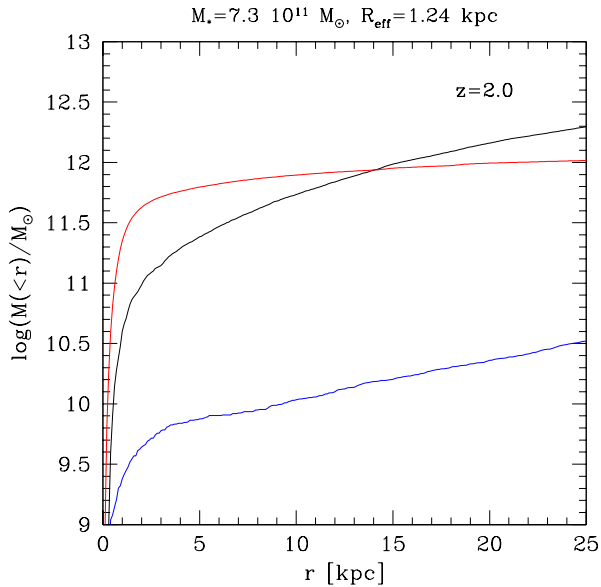}
 \caption{Cumulative mass distribution of gas (blue curve), stars
   (red) and dark matter (black) in a $z=2$ galaxy from the Coma3
   run of $M_*=7.3\times10^{11} M_\odot$ and $R_{\rm{eff}}$=1.24 kpc.} 
\label{fig:mcum}
\end{figure}

To assess, more quantitatively, the effects of mass resolution and
two-body relaxation, we carried out simulations of an individual 
proto-elliptical
galaxy region at much higher mass resolution than used in the 
proto-cluster simulations (cf. Table~\ref{tab:data}). 
The main elliptical galaxy formed has a final $z=0$ stellar mass of about 
$10^{11} M_\odot$, and resides in a ``field'' galaxy region \citep{SL.03}. 
At $z\sim2$ the region contains 5 galaxies of stellar
mass $M_* \sim2 \times10^{10} M_\odot$. Using eq.[11] it is easy to show
that $t_R > t_{H,0}$ and $t_R \gg t_{H,0}$ for simulations ``E4'' and 
``E4HR'', respectively, where $t_{H,0}$ is the present age of the Universe.
In Figures 1-3 are shown $R_{\rm{eff}}$, $\sigma_{*,1D}$ and $\rho_*$
for these five proto-galaxies at $z=2$. Moreover, at $z=0$, the
simulation contains three galaxies of $M_*\sim0.3-1.0\times10^{11} M_\odot$. 
The values of $\rho_*$ for these three galaxies are also shown in 
Fig.~\ref{fig:rhostar}. As can be seen from the figures, the $z\sim2$
high-resolution proto-elliptical results are broadly consistent with
the high force resolution proto-cluster results. This may indicate
that the latter results are not severely affected by 
two-body relaxation effects --- but see below. 
On the other hand, the discrepancy between
the $z=0$ elliptical galaxy and cluster results could indicate that the latter,
at $M_*\la10^{11} M_\odot$, are affected by relaxation
effects. However, there is no fundamental reason why elliptical galaxies in
a ``field'' region and in a cluster regions should have identical 
properties. Hence, in order to determine the characteristics of
$M_*\la10^{11} M_\odot$, $z\sim0$ cluster galaxies, simulations of 
higher mass resolution have to be undertaken - such simulations are in
progress.  

As proto-cluster regions are some of the first regions in which the
growth of small density fluctuations goes non-linear, later causing
re-collapse and galaxy formation,
it is possible that the galaxies formed in the present simulations
on average are even more compact than the limited number of galaxies
observed so far - in fact, this is indicated by Figs.~\ref{fig:reffz2} and
\ref{fig:rhostar}. It is hence important to also simulate other,
and perhaps more representative, regions of the Universe as well. This
will also be the topic of forthcoming work. 

As can be seen from Fig.~\ref{fig:axrat}, the Coma3 and E4HR 
galaxies satisfy $b/a\ga0.4$. Three of the nine galaxies observed
and analyzed by \cite{vD.08} are flatter than this, with 
$b/a = 0.29, 0.30~ \& ~0.35$. This hints at a real discrepancy, which
is only aggravated when considering that the observed axial
ratios are {\it apparent}, such that the observed galaxies may be
intrinsically even flatter.
The reason for the discrepancy is not
clear --- it may originate from comparing galaxies formed in different
environments, but the discrepancy appears to pertain to both
the proto-cluster and proto-elliptical (``field'') simulations.
It may also stem from not comparing observations and simulations in
a consistent way, or, finally, be caused by not including sufficient 
input physics in
the simulations, in this way relating to the ``angular momentum problem''
(e.g., Sommer-Larsen et~al. 2003) - work is in progress to
address this issue as well.\\
Also, the E4HR galaxies appear flatter on average than the Coma3
galaxies of the same stellar mass. This may be an environmental
effect, but may also be due to the $M_*\la10^{11} M_\odot$ galaxies 
spuriously growing rounder with time due to effects of two-body 
relaxation, cf. the discussion above.

In our derivation of expected virial velocity dispersions (eq.[2]),
we have assumed that the galactic mass inside of two effective
radii is dominated by stellar mass. In Fig.~\ref{fig:mcum}, we
show the cumulative distribution of gas, stellar and dark matter
mass for a massive ($M_*=7.3\times10^{11} M_\odot$), $z=2$ galaxy
from the Coma3 run. As can be seen, the inner mass distribution 
is indeed dominated   
by the stellar mass, with the dark and stellar masses only becoming
equal at $r\sim10 R_{\rm{eff}}$. This is a general finding at $z=2$:
for the
mass range for which the mass resolution is definitely sufficient, 
$M_*\ga10^{11} M_\odot$, the fraction of dark matter mass inside of 
$2 R_{\rm{eff}}$ is of the order 20\%. 

Finally, it is clearly of interest, in relation to comparison
to observations, to calculate surface brightness profiles for the
simulated galaxies. A detailed analysis of this, invoking radially
dependent mass-to-light ratios, will be presented in \cite{SLT10}.

\section{Conclusion and outlook}
Motivated by recent observational findings, we
investigate the structure and dynamics of high-mass
($M_*\ga10^{11} M_\odot$) $z\sim2$ galaxies.
Specifically, a number of
cosmological hydrodynamical/N-body simulations of
a $\sim50000$ Mpc$^3$ co-moving (lagrangian), proto-cluster region
have been undertaken. At $z=2$, the highest resolution simulation 
contains $\sim5800$ resolved galaxies, of which 27 have $M_*>10^{11} M_\odot$, 
and 5 have $M_*>4\times10^{11} M_\odot$.\\ 
In addition, we ran a number
of simulations, including one of ultra-high resolution, of a forming
proto-elliptical galaxy.

Total stellar masses, effective radii and characteristic stellar
densities have been determined for all galaxies. At $z=2$, for the 
definitely well resolved (proto-cluster) mass range of 
$M_*\ga10^{11} M_\odot$, we find a
relation of the form $R_{\rm{eff}}=(1.20\pm0.04)M_{*,12}^{1/3}$~kpc,
where $M_{*,12}$ is the total stellar mass in units of $10^{12} M_\odot$.  
This is in line with recent observational findings for very
massive, compact $z\sim2$ galaxies, though somewhat {\it more} compact
than the observed average. The only line-of-sight velocity dispersion
measured for a $z\sim2$ compact galaxy is very large, 
$\sigma_{*,p}=510^{+165}_{-95}$ km/s \citep{vD.09}.   This value can
be matched at about the 1-$\sigma$ level, although a somewhat
larger mass than the estimated $M_*\simeq2\times10^{11} M_\odot$ is indicated.\\
It is found that the massive compact galaxies are strongly baryon dominated
in their inner parts, with typical dark matter mass fractions of order
only 20\% inside of $r=2R_{\rm{eff}}$.

To further study the kinematics and dynamics of the galaxies, we 
calculated axial ratios, rotation velocities and velocity
dispersion tensors for these.
For the above mass range, the galaxies have an average axial ratio 
$<b/a> = 0.64\pm0.02$
with a dispersion of 0.1, and an average rotation to 1D velocity 
dispersion ratio $<v/\sigma> = 0.46\pm0.06$ with a dispersion
of 0.3, and a maximum value of $v/\sigma \simeq 1.1$. 
Rotation and velocity anisotropy both contribute significantly in 
flattening the compact galaxies. Some of the observed compact galaxies
appear flatter than any of the simulated galaxies --- additional work
is required to understand the origin of this possible discrepancy. 

Simulations of different environments, of higher numerical resolution 
and/or based on more detailed input
physics are in progress --- the results will be presented in
forthcoming papers.    

\section*{Acknowledgments}
We are indebted to Peter Johansson for patient help on running and
re-coding the GADGET-2 code. In addition we have benefited from 
discussions with him and Thorsten Naab. We also thank Anders
Sommer-Larsen for assistance in preparing the figures.

We gratefully acknowledge abundant access to the computing facilities
provided by the Danish Centre for Scientific Computing (DCSC). This
work was supported by the DFG Cluster of Excellence ``Origin and Structure
of the Universe''. The Dark Cosmology Centre is funded by the Danish
National Research Foundation. We acknowledge support from the Lundbeck
Foundation.

\label{lastpage}

\end{document}